\documentclass[12pt]{article}
\usepackage{graphicx,amsmath,amssymb,color,bm}
\begin{document}

\centerline {\Large\textbf {Optical Properties of Monolayer Bismuthene in Electric Fields
}}


\centerline{Rong-Bin Chen$^{1,*}$, Der-Jun Jang$^{2}$, Ming-Chieh Lin$^{3,*}$, and Ming-Fa Lin$^{4,*}$}

\centerline{$^{1}$Center of General Studies, National Kaohsiung Marine University, Kaohsiung 811, Taiwan}
\centerline{$^{2}$Department of Physics, National Sun Yat-Sen University, Kaohsiung 804, Taiwan}
\centerline{$^{3}$Department of Electrical and Biomedical Engineering, Hanyang University, Seoul 04763, Korea}
\centerline{$^{4}$Department of Physics, National Cheng Kung University, Tainan 701, Taiwan }

\begin{abstract}

Optical excitations of monolayer bismuthene present the rich and unique absorption spectra. The threshold frequency is not equal to an
indirect energy gap, and it becomes zero under the critical electric field. The frequency, number, intensity and form of the absorption 
structures are dramatically changed by the external field. The prominent peaks and the observable shoulders, respectively, arise from the
constant-energy loop and the band-edge states of parabolic dispersions. These directly reflect the unusual electronic properties, being 
very different from those in monolayer graphene.            
\vskip0.6 truecm


\end{abstract}

\newpage

\bigskip


The layered condensed-matter systems have become one the main-stream materials, as they possess the unusual lattice symmetries and the nano-scaled thickness. Such emergent 2D systems are very
suitable in exploring the diverse physical, chemical, and material fundamental properties, especially for monoelemental graphene \cite{KSNovoselov,KSKim}, silicene \cite{PVogt,LTao}, germanene
\cite{LFLi, MDerivaz}, tinene \cite{FFZhu}, phosphorene \cite{LLi, PYasaei}, antimonene \cite{PAres, JJi} and bismuthene \cite{ZFWang, CSabater, THirahara, THirahara2}.
Recently, few-layer bismuthene could be directly obtained from the mechanical exfoliation, as used for graphene systems in 2004 \cite{KSNovoselov}. Furthermore, they are successfully synthesized on
the 3D Bi$_2$Te$_3$(111)/Bi$_2$Se$_3$(111)/Si(111) substrates \cite{THirahara, THirahara2, FYang, THirahara3}.
The electronic and optical properties are easily tuned by the perpendicular electric field (${F_z\hat z}$), owing to the highly buckled structure. This work is focused on absorption spectra of
monolayer bismuthene in the absence and presence of ${F_z}$.

A rhombohedral bismuth is a well-known semimetal, being similar to the semi-metallic ABC-stacked graphite. However, the intrinsic inetrations are quite different from each other because of the
significant spin-orbital coupling (SOC) in the former. When it becomes a 2D bismuthene \cite{THirahara3, YMKoroteev, JLee, EAkturk} or a 1D quantum wire \cite{JLee, THirahara4}, the semi-metallic
behavior is changed into the semiconducting property. There exist some theoretical and experimental studies on the geometric symmetries \cite{THirahara3, YMKoroteev, JLee, EAkturk}, energy bands
\cite{YMKoroteev, JLee, EAkturk, NFukui}, and transport properties \cite{CSabater}. The buckled structure is clearly identified by the measurements of scanning tunneling microscopy \cite{FYang} and
high-resolution electron diffraction\cite{THirahara3,NFukui}.

For monolayer bismuthene, the tight-binding model and the gradient approximation are, respectively, available in exploring the electronic and optical properties without/with ${F_z\hat z}$. Monolayer
bismuthene consists of the buckled A and B sublattices with a separation of ${l_z=1.84}$ $\AA$, as shown in Fig. 1(a), in which the ${(x,y)}$-plane projections are  the honeycomb lattice with the
primitive unit vectors ${\overrightarrow {a_1}}$ and ${\overrightarrow {a_2}}$ (${|{\overrightarrow {a_1}}|=4.33}$ ${\AA}$). The buckling degree is characterized by the angle (${\theta\,=126^\circ}$)
between the ${Bi-Bi}$ bond and ${z}$-axis (Fig. 1(b)). The low-lying Hamiltonian  is built from the ${sp^3}$ bonding of the ${6s,6p_x,6p_y}$, and ${6p_z}$ orbitals in the presence of significant SOC:

\begin{align}
H& =\underset{\langle i\rangle ,o,s}{\sum }E_{o}C_{ios}^{+}C_{ios}+\underset{%
\langle i,j\rangle ,o,o^{\prime },s}{\sum }\gamma _{oo^{\prime }}^{%
\overrightarrow{R}_{ij}}C_{ios}^{+}C_{jo^{\prime }s}+\underset{\left\langle
\left\langle i,j\right\rangle \right\rangle ,o,o^{\prime },s}{\sum }\gamma
_{oo^{\prime }}^{\overrightarrow{R}_{ij}}C_{ios}^{+}C_{jo^{\prime }s}
\notag \\
& {+}\underset{\langle i\rangle ,p_{\alpha },\text{ }p_{\beta },s,s^{\prime }}{%
\sum }\frac{\lambda _{soc}}{2}C_{ip_{\alpha }s}^{+}C_{ip_{\beta }s^{\prime
}}(-i\epsilon _{\alpha \beta \gamma }\sigma _{ss^{\prime }}^{\gamma })
+l_z\underset{\langle i\rangle ,o,\text{ }s}{\sum }F_{z}C_{ios}^{+}C_{ios}^{{}}\ ,
\end{align}
which ${E_o}$ of the ${6s}$ and ${6p}$ orbitals are, ${-9.643}$ eV and ${-0.263}$ eV, respectively \cite{JHXu}. The second
term is the nearest-neighbor hopping integral ${\gamma _{oo^{\prime }}^{\overrightarrow{R}_{ij}}}$, relying on the type of atomic orbitals, the vector between two atoms (${{\overrightarrow{R}_{ij}}}$), and $\theta$. The various interactions
in the ${sp^3}$ bondings are ${V_{pp\pi} = -0.679}$ eV,
${V_{pp\sigma}}$ = 2.271 eV, ${V_{sp\sigma}}$= 1.3 eV, and ${V_{ss\sigma}= -0.703}$ eV, as
clearly indicated in Fig. 1(c).  The next-nearest-neighbor hopping integrals in the third term, being  independent of $\theta$, cover
${V_{pp\pi}=0.00 4}$ eV, ${V_{pp\sigma}= 0.303}$ eV, ${V_{sp\sigma} = 0.065}$ eV, and
${V_{ss\sigma}= -0.007}$ eV. The last term stands for the intra-atomic SOC, ${V_{soc}=\lambda_{soc}\,{\overrightarrow L\cdot\,\overrightarrow s}}$ (${\lambda_{soc}=1.5}$ eV). ${\alpha\,}$, ${\beta\,}$, and ${\gamma}$ denote the ${x}$, ${y}$ or ${z}$ coordinates, and $\sigma$ is the Pauli spin matrix.  It could alter the spin configurations between the ${6p_z}$ and ${(6p_x,6p_y)}$ orbitals $\&$ the
${6p_x}$ and ${6p_y}$ orbitals. The final term is the Coulomb potential energy due to an electric field $F_z$.                                                                                          
When monolayer bismuthene exists in an electromagnetic wave, the occupied valence states are excited to
the unoccupied conduction ones with the same wave vector. The electric dipole moment will induce the  vertical optical excitations, in which the spectral intensity under the Fermi golden rule is
characterized by

\begin{align}
{\ }A(\omega )& \propto \underset{v,c,n,n^{\prime }}{\sum }%
\int_{1^{st}BZ}\frac{dk_{x}dk_{y}}{(2\pi )^{2}}Im[\frac{1}{E_{n^{\prime }}^{c}(k_{x},k_{y})-E_{n}^{v}(k_{x},k_{y})-(\omega \mathbf{+}i\gamma )}]
\notag \\
& \times \left\vert \left\langle \Psi
_{n^{\prime }}^{c}(k_{x},k_{y})\right\vert \frac{\widehat{E}\cdot
\overrightarrow {P}}{m_{e}}\left\vert \Psi _{n}^{v}(k_{x},k_{y})\right\rangle
\right\vert ^{2} ,
\end{align}%
where ${\gamma}$ (=10 meV) is the broadening parameter due to various  deexcitations and $n$ represents the number of energy subbands, measured from the Fermi level. The electric polarization is in the
${\hat x}$-/${\hat y}$-direction. The absorption spectrum depends on the joint density of states (JDOS) between the valence and conduction states and the square of the velocity matrix element. The
former is associated with energy dispersions or DOS while the latter is evaluated from the gradient approximation, being successfully
utilized in graphene-related systems \cite{ZQLi, LMZhang, ABKuzmenko, YHLai}.

Monolayer bismuthene possesses three low-lying energy bands (Fig. 2(a)), with two valence and one conduction bands centered about the $\Gamma$ point. The first conduction band ($c_1$) has
parabolic dispersions near the Fermi level along any directions, as observed in the second valence band ($v_2$).   A sombrero-shaped energy dispersion, with the constant-energy loop, is revealed
in the first valence band ($v_1$) with the local extremum deviating from $\Gamma$, so it could also be regarded as the 1D parabolic band. This is responsible for an indirect energy gap of
${E_g^i=0.295}$ eV. The double spin degeneracy is clearly destroyed by an electric field under the $z$-plane mirror asymmetry. The spin-up- and spin-down-dominated energy bands are similar to
each other, in which the extremal points are situated at/near the $\Gamma$ point. With the increase of $F_z$, the conduction and valence bands approach to the Fermi level until the critical
field strength of ${F_{zc}=0.82}$ V/$\AA$ and then deviate from it. Energy gap diminishes to zero and opens again during the variation of $F_z$.
Specifically, the linear Dirac-cone structures  could be formed at ${F_{zc}}$, while the normal parabolic bands appear beyond it.

DOS, as shown in Fig. 2(b), possess the special structures due to the van Hove singularities. The shoulder structures, the square-root asymmetric peaks, and the V-shaped form, respectively,
arise from the local extremes, the constant-energy loops, and the linear Dirac cones in the energy-wave-vector space. The minimum/minima of the first conduction band  exhibits an obvious step
structure in the absence and presence of $F_z$. Apparently, its energy is very sensitive to the strength of $F_z$. It should be noticed that a step structure in the $v_2$ DOS becomes an
asymmetric peak at sufficiently high field  (${F_z\ge\,0.4}$ V/$\AA$).  However,  one prominent peak of the $v_1$ DOS is changed into two peaks/the V-form/the step structures
before/equal/beyond the critical field strength (${F_{zc}=0.82}$ V/$\AA$), in which the higher-energy one is stronger compared with the lower-energy one. The predicted low-lying special
structures in DOS could be directly verified by scanning tunneling microscopy \cite{YNiimi, JAStroscio}. That will be directly reflected in optical absorption spectra.
Monolayer bismuthene presents the unusual optical excitations (Fig. 3), being easily tuned by the external electric field in terms of the frequency, number, and intensity of special structures.
There are  prominent  asymmetric peaks and shoulders arising from the JDOS. Without $F_z$, the threshold frequency is characterized by a strong asymmetric absorption peak (the red curve), so
that the optical gap (${E_g^o=0.35}$ eV) is different from the indirect band gap (0.295 eV). Specifically, it comes from constant-energy loop in the $v_1$ band and the corresponding $c_1$
conduction states with the same wave vectors. Moreover, an obvious shoulder appears at higher frequency of ${\omega_{s}\sim\,0.52-0.53}$ eV. The initial peak is split into two peaks with the
lower intensities as $F_z$ gradually grows (the blue, black, and black-squared blue curves), especially for the lowest threshold one. The splitting spectrum is accompanied with the decrease of
optical gap, in which the vanishing threshold frequency and the featureless spectrum could occur at the critical field. Beyond ${F_{zc}}$, the initial peak structures become a  shoulder structure
(the pink-triangled black curve). On the other hand, absorption frequency of shoulder structure might decline. This structure could be merged with the prominent peak and it would disappear near
${F_{zc}}$, or a shoulder structure is recovered beyond ${F_{zc}}$. The absence of absorption structure is revealed  in the range of ${0.6}$ V/$\AA$${\le\,F_z\le\,F_{zc}}$ .

     The spectral intensity strongly depends on the strength of electric field, being very useful in fully understanding the absorption features and the comparison with the experimental measurements.
The optical threshold  frequency is determined by the lowest  absorption boundary. $E_g^o$ nonlinearly declines with the increase of $F_z$, becomes zero at ${F_{zc}}$, and then gradually increases in the further increment of $F_z$. Apparently, the second absorption structure split from the first one at ${F_z=0}$ gradually approaches to the third one. Both of them  are merged together at ${F_z=0.52}$ V/$\AA$, and the merged structure disappears at ${F_{zc}}$ Finally, a new absorption structure occurs at the higher frequency beyond ${F_{zc}}$. In short, monolayer bismuthene might possess three, two, one and zero absorption structures, and they are easily tuned by the electric field because of the significant buckled geometry.

The above-mentioned features of absorption spectra could be examined by various optical spectroscopies, such as the absorption, transmission, reflection, Raman scattering and Rayleigh scattering
spectroscopies. They are successfully utilized to identify the low-frequency optical excitations in AB- and ABC-stacked trilayer graphenes \cite{CLLu, CWChiu, CYLin}. For the former, the special
shoulder structure at ${\omega\,\sim\,0.5}$ eV, which is induced by the band-edge state transitions from the parabolic valence band to the similar conduction band of another pair (not the same pair),
is identified by the infrared reflection and absorption spectroscopies \cite{ZQLi, LMZhang, ABKuzmenko}. Recently, the similar optical measurements have verified the low-frequency optical properties, showing a clear evidence of two characteristic peaks associated with the surface-localized flat and sombrero-shaped energy bands \cite{KFMak}. The experimental examinations on the predicted
low-frequency optical excitations will be very useful in exploring the effects due to the geometric symmetry, the intrinsic interactions, the spin-orbital coupling, and the external field.
Monolayer bismuthene exhibits the unusual electronic and optical properties, being in sharp contrast to the linear Dirac-cone structure near the K/K$^\prime$ point without the featured absorption
spectrum in monolayer graphene \cite{DSLAbergel}. The low-lying $c_1$, $v_1$ and $v_2$ bands are centered about/near the $\Gamma$ point. The former two have an indirect gap of ${\sim\,0.295}$ eV but
an optical gap of ${\sim\,0.35}$ eV.  They show the threshold asymmetric peak and the higher-frequency shoulder, corresponding to ${v_1\to\,c_1}$ and ${v_2\to\,c_1}$  associated with the
constant-energy loop and the parabolic extremum, respectively. The initial strong peak becomes two ones as $F_z$ gradually increases. The step structure might be changed into the
observable peak at sufficiently large $F_z$. The frequency, number, intensity, and form of optical spectra dramatically alter during the variation of $F_z$.

\bigskip
\bigskip

\centerline {\textbf {ACKNOWLEDGMENT}}%

\bigskip
\bigskip

We thank Feng-Chuan Chuang for helpful discussions. This work is supported by the Ministry of Science and Technology of Taiwan, under grant Nos. MOST 105-2112-M-006-007-MY3, MOST-104-2112-M-110-002-MY3, and the support under NSYSU-NKMU Joint Research
Project Nos. 105-P005 and 106-P005.

\newpage

\begin{figure}
\centering
\includegraphics[width=0.9\textwidth]{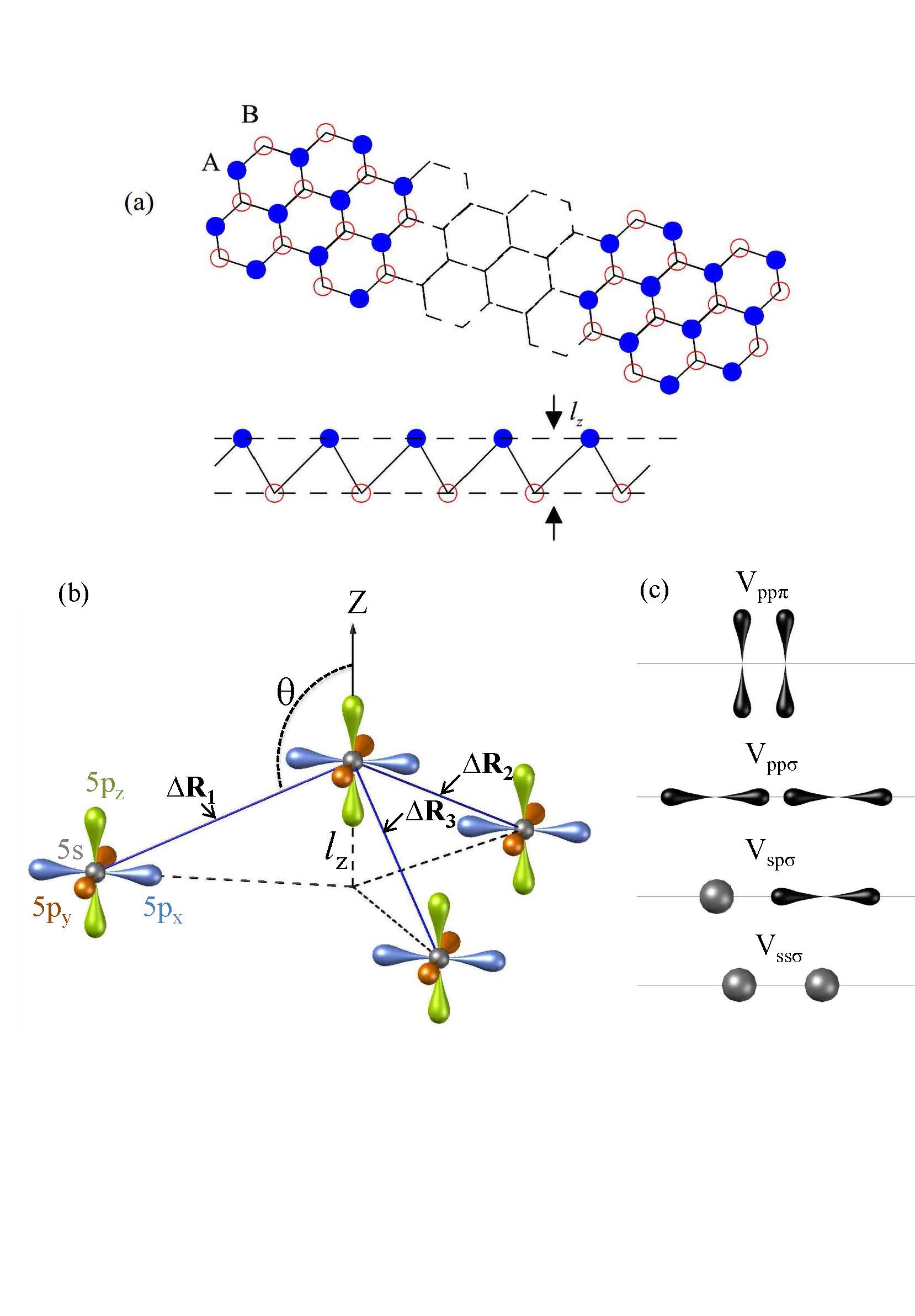}
\caption{(a) the top and side views of monolayer bismuthene, (b) the ${sp^3}$ orbital hybridizations, and (c) the various $\pi$ and $\sigma$ bondings.}
\label{FIG:1}
\end{figure}

\begin{figure}
\centering
\includegraphics[width=0.9\textwidth]{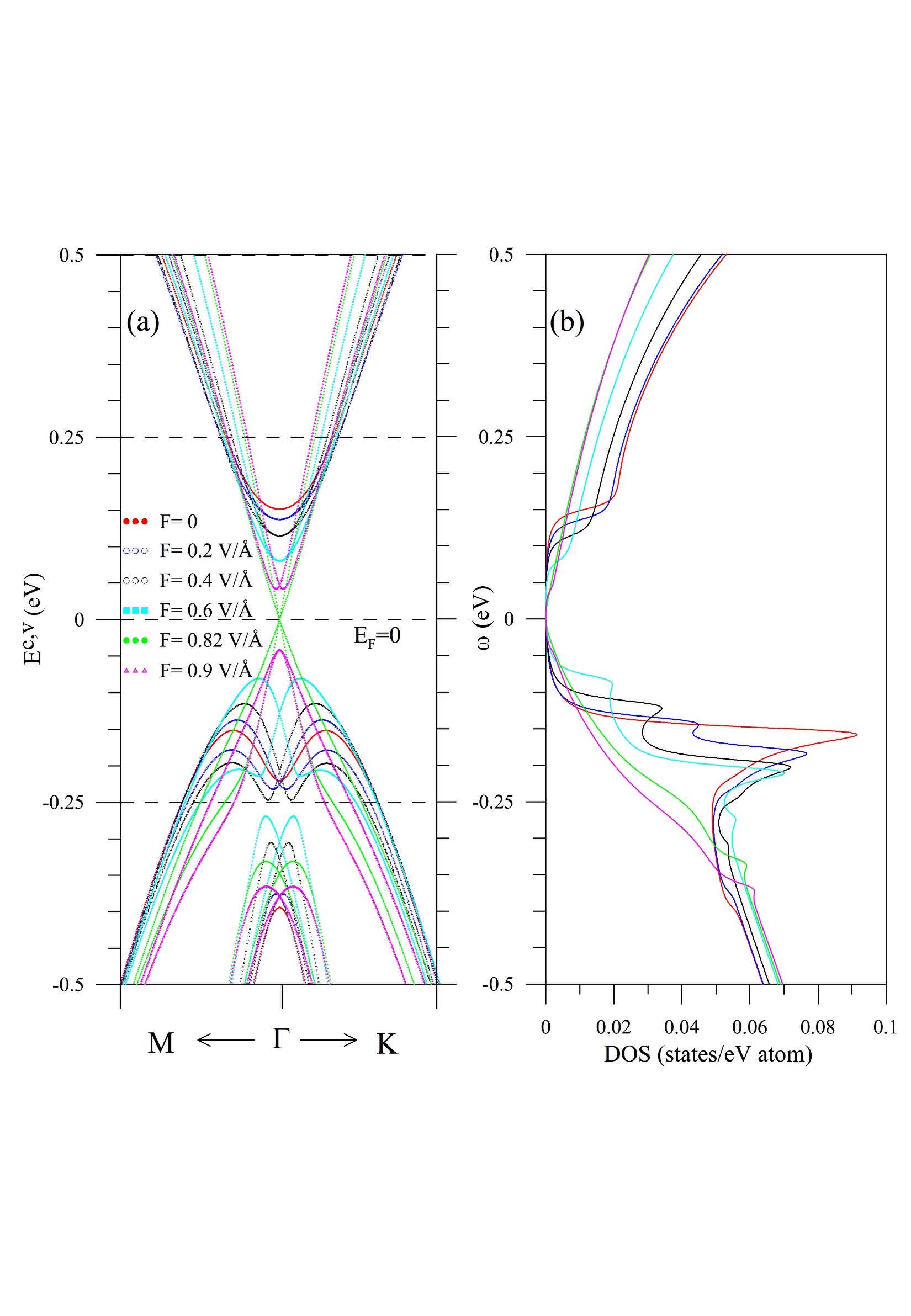}
\caption{The low-lyling (a) band structures and (b) densities of states of monolayer bismuthene at distinct electric fields.}
\label{FIG:2}
\end{figure}

\begin{figure}
\centering
\includegraphics[width=0.9\textwidth]{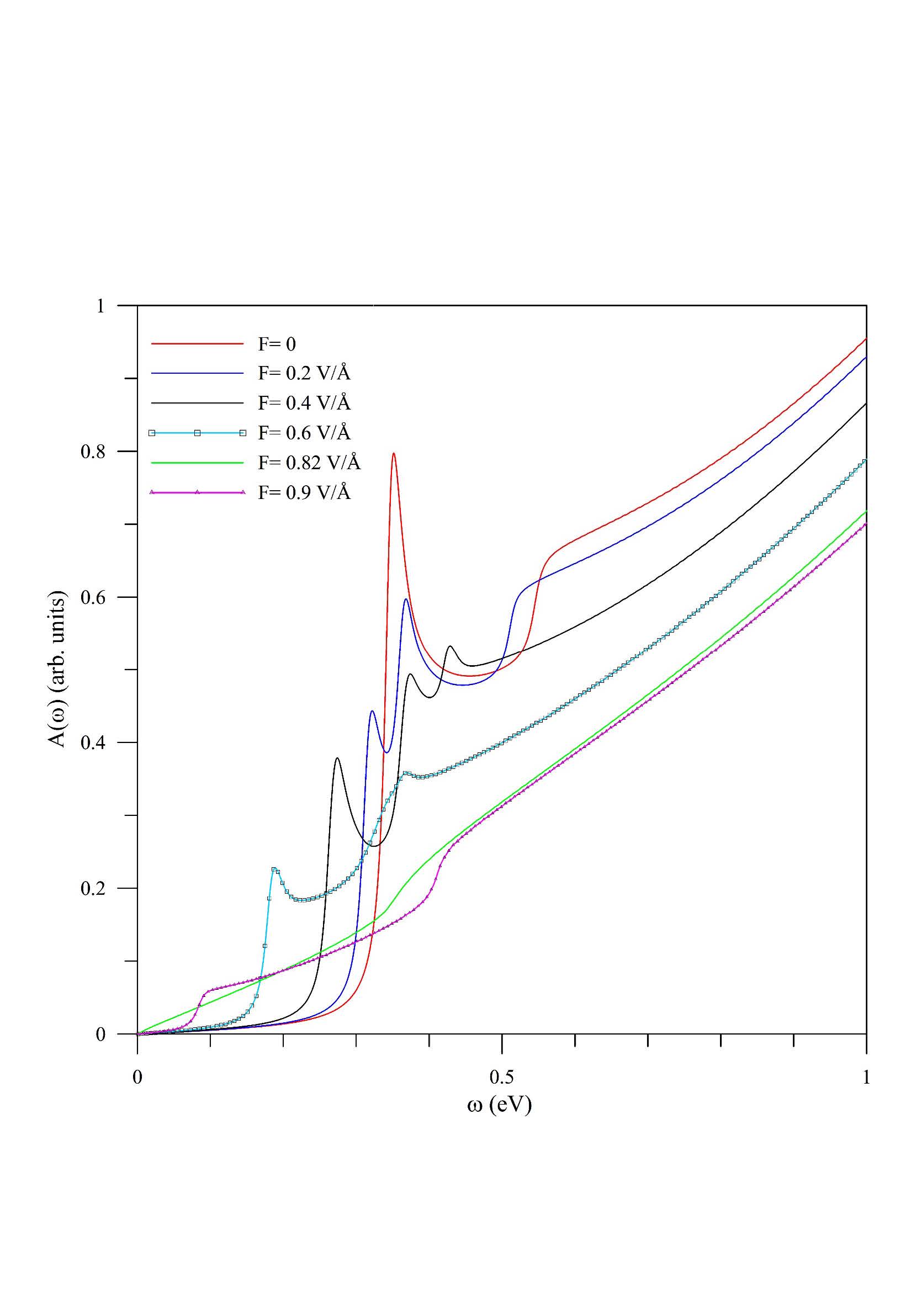}
\caption{The optical absorption spectra of monolayer bismuthene under the various electric fields.}
\label{FIG:3}
\end{figure}

\begin{figure}
\centering
\includegraphics[width=0.9\textwidth]{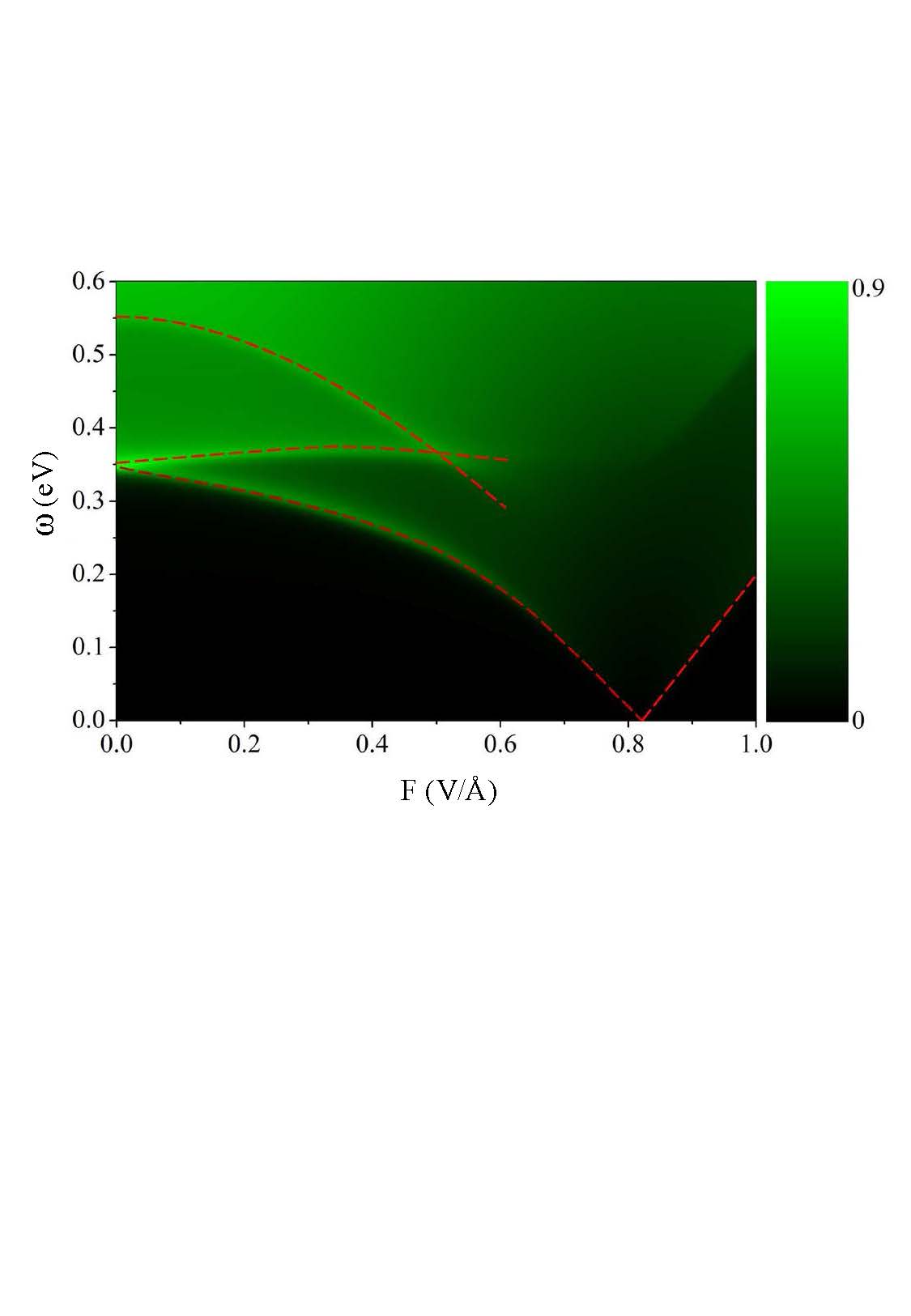}
\caption{The electric-field-dependent spectral intensities, with the absorption structures indicated by the red dashed curves.}
\label{FIG:4}
\end{figure}

\end{document}